\documentclass[epj,final]{svjour}
\usepackage{graphics}
\usepackage{amssymb,amsmath}
\usepackage{array}
\usepackage{dcolumn}

\begin{document}

\title{$S\!r_{14}C\!u_{24}O_{41}$ : a complete model for the
chain sub-system}

%
%
%
%
%


\author{Alain Gell\'e\inst{1} \and Marie-Bernadette Lepetit\inst{1,2}}
\institute{Laboratoire de Physique Quantique, IRSAMC~-~CNRS~UMR~5626,
Universit\'e Paul Sabatier, 118 route de Narbonne, F-31062 Toulouse
Cedex 4, FRANCE \and Laboratoire CRISMAT, ENSICAEN~-~CNRS~UMR~6508, 6
boulevard Mar\'echal Juin, F-14050 Caen Cedex, France
\\ \email{lepetit@ensicaen.fr}}

\abstract{ A second neighbor $t-J+V$ model for the chain subsystem of
the $S\!r_{14}C\!u_{24}O_{41}$ has been extracted from ab-initio
calculations. This model does not use periodic approximation but
describes the entire chain through the use of the four-dimensional
crystallographic description. Second neighbors interactions are found
to be of same order than the first neighbors ones. The computed values
of the second neighbors magnetic interaction are coherent with
experimental estimations of the intra-dimer magnetic interactions,
even if slightly smaller. The reasons of this underestimation are
detailed. The computed model allowed us to understand the origin of
the chain dimerisation and predicts correctly the relative occurrence
of dimers and free spins. The orbitals respectively supporting the
magnetic electrons and the holes have been found to be essentially
supported by the copper $3d$ orbitals (spins) and the surrounding
oxygen $2p$ orbitals (holes), thus giving a strong footing to the
existence of Zhang-Rice singlets.}

\PACS{71.10.Fd, 71.27.+a, 71.23.Ft}

\date{\today}

\maketitle

\section{Introduction}

One-dimensional quantum systems have attracted a lot of attention in
the past decade due to the large diversity of their low energy
physics. In particular, spin-chains and spin-ladders systems have been
extensively studied. The characteristic of the
$S\!r_{14-x}A_xC\!u_{24}O_{41}$ ($A=C\!a, B\!a,Y,B\!i$, etc.) family
of transition-metal oxides is that they are composed of both
spin-chains and spin-ladders sub-systems. These compounds are formed of
alternated layers (in the $(a,c)$ plane) of each of the two
subsystems~\cite{struc1}. Both ladders and chains are in the ${\bf c}$
direction, however their respective translation vector ($\bf c_{c}$
and ${\bf c_l}$) are incommensurate. In the pure compound,
$S\!r_{14}C\!u_{24}O_{41}$, the layers are largely separated ($\simeq
3.3$ \AA) and considered as electronically non
interacting. Nevertheless, the low energy properties of
$S\!r_{14}C\!u_{24}O_{41}$ agree neither with those of spin chains,
nor with the properties of spin ladders.

$S\!r_{14}C\!u_{24}O_{41}$ is a semiconductor with a $0.18\, \rm eV$
gap~\cite{Magn96B} at $T<T^\star=250\, K$. The spin ladders have a
singlet ground state with a spin gap of about $35-47\,\rm
meV$~\cite{Neut96,RMN98,RMN97}. Surprisingly the spin chains also
exhibit a singlet ground state with a spin gap of $11-12\, \rm
meV$~\cite{Magn96B,RMN97,RMN98B,Neut98,Thermo00} while homogeneous
spin chains are known to be gap-less in the
spin channel. Susceptibility and ESR measurement~\cite{Magn96,Magn96B}
suggested that the spin gap in the chains is due to the formation of
weakly interacting spin dimers. Neutron scattering
experiments~\cite{Neut96} have later confirmed their existence.

The spins are supported in the chain subsystem by the $3d_{ac}$ orbitals
of the $C\!u^{2+}$ ions, and in the ladder subsystem by the $3d_{a^2-c^2}$
orbitals of the $C\!u^{2+}$ ions. In the chain subsystem the magnetic
orbitals are coupled via two nearly $90^\circ$ $C\!u$--$O$--$C\!u$
bonds. Let us note at this point that, in such geometries,
the super-exchange paths through the oxygen orbitals interact destructively
and therefore nearest-neighbor (NN) exchange interactions are
expected to be small and ferromagnetic. In the ladder subsystem, the
picture of the NN interactions is very different since there are
mediated via nearly $180^\circ$ $C\!u$--$O$--$C\!u$ angles. Such
geometries are known to produce strong super-exchange mechanism via
the bridging ligands and thus large anti-ferro\-magnetic interactions.

Formal charge analysis shows that the $S\!r_{14}C\!u_{24}O_{41}$
compound is intrinsically doped with six holes by formula unit
(f.u.). Similar to high-$T_c$ superconductors, the holes were expected
to be mainly supported by the oxygen $2p$ orbitals and to form
Zhang-Rice~\cite{ZR} singlets with the associated-copper hole. NEXAFS
(Near Edge X-ray Absorption Fine Structure) experiments~\cite{XRay00}
later supported this assumption.

It has been established, from neutron scattering~\cite{Neut96} and X
ray spectroscopy~\cite{XRAY98} experiments, that the chain dimeric
units are formed by two next-nearest-neighbor (NNN) spins separated by a  
hole. $C\!u$ NMR
measurements exhibited the presence of two kinds of holes on the
chains~\cite{RMN98}, namely with intra- and inter-dimer
localization. The relative occurrence of the two types of holes 
($0.65\simeq 1/2$) led the authors to propose a charge-order model
with dimers separated by two holes. This assumption has been confirmed
by neutron scattering experiments~\cite{Neut98,Neut99} that have shown
to be consistent with a five units periodicity. Such a picture leads
to a chain filling of 6 holes per f.u., that is with all the holes
located on the chains. 

The question of the holes repartition between the chain and ladder
subsystems is however still under debate. Indeed, NEXAFS
experiments~\cite{XRay00} evaluated to $0.8$ the number of holes on
the ladders legs oxygens. Magnetic susceptibility
measurements~\cite{Magn96B} exhibit a filling of $3.5$ spins per f.u.,
that is $0.5$ more holes than the maximum number given by the formal
charge analysis. Finally, magnon-hole scattering
experiments~\cite{Magnon04} suggest a total localization of the holes
on the chain subsystem for temperatures lower than $T^*\simeq 200\, K$.

The origin of the chain electronic dimerisation has only recently
being elucidated. The hypothesis of a spin-Peierls transition has been
rapidly eliminated since the expected signatures in the magnetic
susceptibility and specific heat were not found. It thus has long been
supposed that the dimerisation originates from the competition between
first and second-neighbor spin interactions as predicted by Majumdar
and Gosh~\cite{gosh} and Haldane~\cite{hald}.  We recently showed,
using ab initio calculation, that the origin of the dimerized state is
of structural origin, even if not of spin-Peierls
one~\cite{SrCa}. Indeed, this is the structural incommensurate
modulations of the chain subsystem, with the periodicity of the ladder
one, that strongly modulate the spin/hole orbital energy and localize
the spins in a dimer pattern.

These incommensurate structural modulations, that are most of the time
neglected, have thus proved to be crucial for the low energy
properties of this system. Independently to the on-site orbital energy
modulations, the structural distortions of the chain subsystem were
shown to be responsible for strong modulation of the first neighbor
effective interactions~\cite{SrCa}, namely NN hopping and NN exchange
integrals. One can thus suppose that these structural modulations will
also affect the second neighbor interactions that are expected to be
of the same order of magnitude than the NN ones in these $90^\circ$
oxygen bridged systems. Indeed, while the super-exchange path through
the bridging oxygens is hindered in the NN interactions, this is not
the case for NNN ones where interaction paths through two oxygen
bridges are favored.  In addition, these second neighbor effective
integrals are crucial for the low energy properties of this system
since the structural modulations localize the spins into NNN 
dimers.

The present work thus aims at giving a complete description of the
chain subsystem within a second neighbor $t-J+V$ model, taking
explicitly into account the structural incommensurate modulations.
The incommensurate character of the problem will be treated within the
four-dimensional representation and we will describe the electronic
structure without the use of a periodic approximation. Using this
complete model, we will further analyze our results as a function of
the four crystallographic coordinate and in particular study the
different (quasi)independent spin entities, their relative proportion
and positions. This four dimensional method presents the advantage of
giving a complete description of these incommensurate systems, without
the use of any periodic approximation.

The next section will describe the ab-initio methods used in the
present calculations. Section III will analyze the nature of the holes
and yield ab-initio evidence of Zhang-Rice singlets. Section IV will
give the second neighbor $t-J+V$ model as a function of the fourth
crystallographic coordinate $\tau$. Section V will details the NNN
double-bridged super-exchange mechanism and its calculations. Section
VI will analyze the exchange fluctuations along the chains. Finally
the last section will conclude.

\section{The ab-initio method} 

It is well known that magnetic and transfer interactions are
essentially local in strongly correlated systems and can thus be
accurately evaluated using embedded fragment ab-initio spectroscopy
methods~\cite{revue}. The long range electrostatic effects are treated
within a bath composed of total-ion pseudo-potentials~\cite{TIPS} and
charges. The open-shell character of the magnetic/hole orbitals, the
strong electronic correlation as well as the screening effects are
efficiently treated using quantum-chemistry ab-initio spectroscopy
methods. The present calculations were performed using the
Difference-Dedicated Configuration Interaction method~\cite{DDCI}
(DDCI). This configuration interaction method has been specifically
developed in order to accurately calculate the low energy states
(spectrum and wave functions) of open-shell, strongly correlated
formally finite systems. Used on embedded crystal fragments, the DDCI
method allow to obtain effective local interactions with a great
accuracy and reliability. For instance, it was proved to be very
efficient on copper and vanadium compounds such as high $T_c$ copper
oxides~\cite{DDCIhtc} or the famous $\alpha^\prime N\!aV_2O_5$
compound~\cite{vana}.

\begin{figure}[h] 
 \center
\resizebox{8cm}{!}{\includegraphics{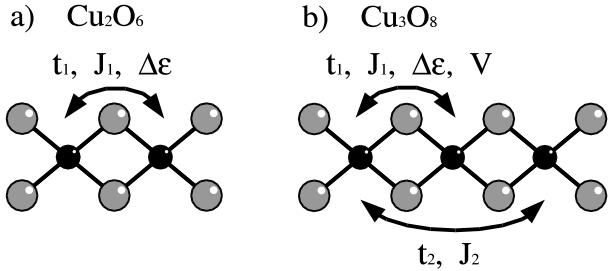}}
\caption{a) Schematic representation of the computed fragments. a) two
centers b) three centers. The gray circles represent oxygen atoms,
while the black circles represent the copper atoms.}
\label{frag}
\end{figure}

The quantum fragments are defined so that to include (i) the magnetic
centers, (ii) the bridging oxygens mediating the interactions, and
(iii) the first coordination shell of the preceding atoms which is
responsible for the essential part of the screening effects.
Second-neighbor interactions are thus extracted using three centers
$C\!u_3 O_8$ fragments (see figure~\ref{frag}b).  NNN exchange
interactions $J_2$ are obtained from the doublet-quartet excitations
energies when three magnetic electrons are considered.  NNN hopping
and first neighbor bi-electronic Coulomb repulsion are obtained from
the 3 singlets and 3 triplets of the fragment containing one magnetic
electron less.  Let us point out that the 3 centers calculations also
yield the NN interactions, which were previously obtained using two
centers $C\!u_2 O_6$ fragments~\cite{SrCa}. The comparison between the
evaluations of the first-neighbor integrals obtained from the 2
centers and 3 centers fragments allows us to verify the relevance of
the chosen model and the fragment size dependence of our calculations.

A least-square fit method is used in order to extract the effective
parameters from the ab-initio calculations.  The conditions imposed
for this purpose are that the effective model should reproduce
\begin{itemize} \itemsep +0.8ex
\item the computed excitation energies, 
\item the projection of the computed wave-functions within the
configuration space generated by the magnetic orbitals.
\end{itemize}
Let us note that the norm of the later projections give us a measure
of the model validity. Indeed, in the present work, the ab-initio
wave-functions (which are expanded over several millions of
configurations) project over the model configuration space based only
on the magnetic orbitals (typically of the order of 10
configurations), with a norm as large as $0.8$. One can thus assume
that this model space is appropriate to describe the low energy
physics of the system.

Let us now address the embedding problem for incommensurate
systems. The usual embedding technique consists of reproducing the
Madelung potential using a set of point charges, located at the
crystallographic positions over a box of at least 15 to $20\,\rm \AA$
around the fragment~\cite{struct}. These charges are adjusted on the
bath borders according to an Evjen procedure~\cite{evjen}. In the case
of a periodic ionic crystal this method insures the nullity of both
the system charge and dipole moment. These two conditions are
necessary to insure a good representation of the Madelung
potential. In the present incommensurate case, however, the Evjen
procedure fails to suppress the dipole moment. This is due to the fact
that the chain and ladder subsystems are not electrostatically
neutral. The relative displacement of one compared to the other thus
induces a dipole moment in the chain/ladder direction. In order to
solve this problem, and to cancel out the dipole moment contribution,
we adjusted the charges of the outermost unit cells of each subsystem,
in the $c$ direction, using a global scaling factor for each adjusted
cell~\cite{env2}.

The calculations have been done on 11 equivalent fragments located at
11 successive positions in the chain direction. These 11 fragments
give a quite good representation of the different distortions
occurring on the chain subsystem. In order to fully represent the
whole chain subsystem, these 11 sets of results have been
extrapolated, using Fourier's series analysis, as a function of the
crystallographic fourth coordinate $\tau$, which is associated with
the system incommensurate modulations.

Let us notice that in a complete crystallographic
description~\cite{struct4}, each atom possesses a fourth fractional
coordinate $\tau_i = {\bf r_i \cdot k} =z_i c_c/c_l$, where $\bf r_i$
is the atom position in the average structure, ${\bf k}= {\bf c_c^\star} c_c/c_l$ is the
modulation vector, $z_i$ is the fractional coordinate of the atom in
the ${\bf c}$ direction. In the model Hamiltonian used in the present work,
$\tau$ corresponds to the fourth coordinate of the chain unit-cell
copper atom. It is defined except for a constant.

\section{Nature of the holes}

As previously noticed, it has been supposed in the literature that
the holes (both in the chain and ladder subsystems) are not located on
the copper atoms but rather on the surrounding oxygens. This assumption
has been done by analogy with the high $T_c$ copper oxides and has
later been comforted by NEXAFS experiments~\cite{XRay00} as far as
ladder holes are concerned. We have thus derived from our ab-initio
calculations the nature and composition of both the magnetic
(supporting the spins) and hole orbitals.

The magnetic orbitals have been obtained, from the 2 centers fragment
calculations, as the triplet natural orbitals (eigenfunctions of the
one-electron density matrix) with an occupation number close to 1. In
order to locate the hole orbitals we compared two calculations on the
same fragment with one electron difference. The hole orbitals have
been extracted from the difference between twice the triplet
density-matrix and the sum of the two doublet ones. The hole orbitals
are thus the two eigenfunctions of the resulting matrix associated
with eigenvalues close to 1.

\begin{figure}[h] 
 \center
\resizebox{8cm}{!}{\includegraphics{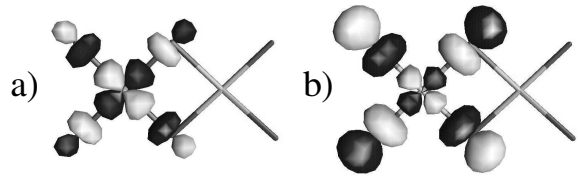}}
\caption{a) Example of magnetic orbital supporting the spin, b)
associated hole orbital.}
\label{f:orb}
\end{figure}

Figure~\ref{f:orb} represents a typical example of both the magnetic
(a) and hole (b) orbitals on a site. It clearly appears that while the
magnetic orbital is essentially supported by the $3d_{ac}$ copper
orbital with a delocalization tail on the surrounding $2p$ oxygens 
orbitals, the hole orbital is essentially supported by the $2p$
oxygen orbitals with a small tail on the $3d_{ac}$ orbital of the copper
atom. The average repartition between copper and oxygens ---~over the
11 calculations~--- are as follow :
\begin{itemize} \itemsep 1.2ex
\item for the magnetic orbital~: $67\%$ on the copper and $33\%$ on the
oxygens, 
\item for the hole orbital~: $15\%$ on the copper and $85\%$ on the
oxygens.
\end{itemize}
The fluctuation of the above repartition on the 11 computed fragments
is very small with a standard deviation of only $1.3\%$. This
stability of the copper-oxygen composition of the hole and spin
supporting orbitals is quite noticeable since the electrostatic
potential modulation on both the copper and oxygen atoms are very
strong, with an amplitude of $4\rm eV$.  Finally, we would like to
point out the stronger oxygen character of the hole orbitals in this
compound, compared to what was found by Calzado {\it et
al}~\cite{AbI_La2CuO4} in the high $T_c$ parent compound
$L\!a_2C\!uO_4$. Indeed, they found only a $50\ \%$ oxygen character
of the hole supporting orbital.

The present results show the existence of different orbitals,
respectively supporting the spins and the holes, and constitute a
direct evidence of the ZR singlets in the system.

\section{Second neighbor $t-J+V$ model}

The parameters of a second neighbor $t-J+V$ model (see
eq.~\ref{eq:tj}) were extracted from the ab initio calculations on the
eleven positions of the fragments along the chain. 
\begin{eqnarray} \label{eq:tj}
H_{t-J+V} &=&  \quad \sum_{i} {\it \varepsilon(\tau_i)}\; n_i  \nonumber\\
&& + \sum_{i} {\it t_1(\tau_i)} \sum_\sigma 
\left( a^\dagger_{i, \sigma} a_{i+1, \sigma} + a^\dagger_{i+1, \sigma} a_{i, \sigma} \right)  \nonumber\\
&& - \sum_{i} {\it J_1(\tau_i)} 
\left({\bf S_{i} S_{i+1}} -1/4\; n_i n_{i+1} \right)  \nonumber\\ 
&& + \sum_{i} {\it t_2(\tau_i)} \sum_\sigma 
\left( a^\dagger_{i, \sigma} a_{i+2, \sigma} + a^\dagger_{i+2, \sigma} a_{i, \sigma} \right)  \nonumber\\
&&- \sum_{i} {\it J_2(\tau_i)} 
\left({\bf S_{i} S_{i+2}} -1/4\; n_i n_{i+2} \right) \nonumber\\
&&+ \sum_{i} {\it V(\tau_i)}\; n_in_{i+1} 
\end{eqnarray}
where $\tau_i$ is the fourth coordinate of site $i$. $a^\dagger_{i,
\sigma}$, $a_{i, \sigma}$ and $n_i$ are the usual creation,
annihilation and number operators, $\sigma$ is the spin quantum
number.

We thus obtained eleven sets of parameters that were further fitted as
a function of the fourth crystallographic coordinate $\tau_i$.
The fit was done using Fourier series, according
to the following expression
\begin{eqnarray} \label{eq:fit}
&& a_0 + \sum_n a_n \cos{\left(2\pi n \tau - \varphi_n \right)} 
\end{eqnarray}
Only terms with a non negligible contribution to the series have been
retained. It results that the orbital energies, hopping integrals and
NN repulsion terms can accurately be obtained from a unique
cosine. Exchange integrals, however, necessitate two components to be
reliably reproduced, as can be expected from the quadratic dependence
of the latter to the hopping integrals.

The results are summarized in table~\ref{t:fits}. Let us notice that
only terms with even frequencies have a non negligible contribution  
 ($n=2$ and $n=4$ in eq.~\ref{eq:fit}). This can be interpreted 
as a doubling of the modulation vector ${\bf k}$. 
The fourth coordinate of a unit cell  is thus given by 
$$\tau = z \times 2 \, c_c/c_l \simeq z \times 2 \times 7/10 $$ where
$z$ is the fractional coordinate in the ${\bf c}$ direction of the
unit cell copper atom. It thus clearly appears that the model
Hamiltonian presents a 5 unit cells quasi-periodicity.  This point is
in agreement with the neutron scattering
experiments~\cite{Neut98,Neut99} that see a pseudo periodicity of the
spin arrangement corresponding to five chain unit cells.

\begin{table}[h]
\center
\begin{tabular}{l|rrrrrr}
meV  & \multicolumn{1}{c}{$\varepsilon$}  & \multicolumn{1}{c}{$V$} & \multicolumn{1}{c}{$t_1$ } & \multicolumn{1}{c}{$t_2$} & \multicolumn{1}{c}{$J_1$} & \multicolumn{1}{c}{$J_2$} \\ \hline 
$a_0$ & 0 & 661 & 132.0 & 214.3 & 20.88 & -6.81 \\
$a_2$ & 600 & -63 & -67.2 & -45.3 & -2.63 & 1.80 \\
$a_4$ & & & & 4.2 & -2.29 & 0.26 \\[1ex]
$\varphi_2$ & 0 & -0.353 & -0.401 & -0.442 & -0.329 & -0.450 \\
$\varphi_4$ & & & & 0.521 & -0.411 & -0.368
\end{tabular}
\caption{Analytic fit of the $t-J+V$ second neighbor model.}
\label{t:fits}
\end{table}

\begin{figure}[h] 
\center
\resizebox{7cm}{!}{\includegraphics{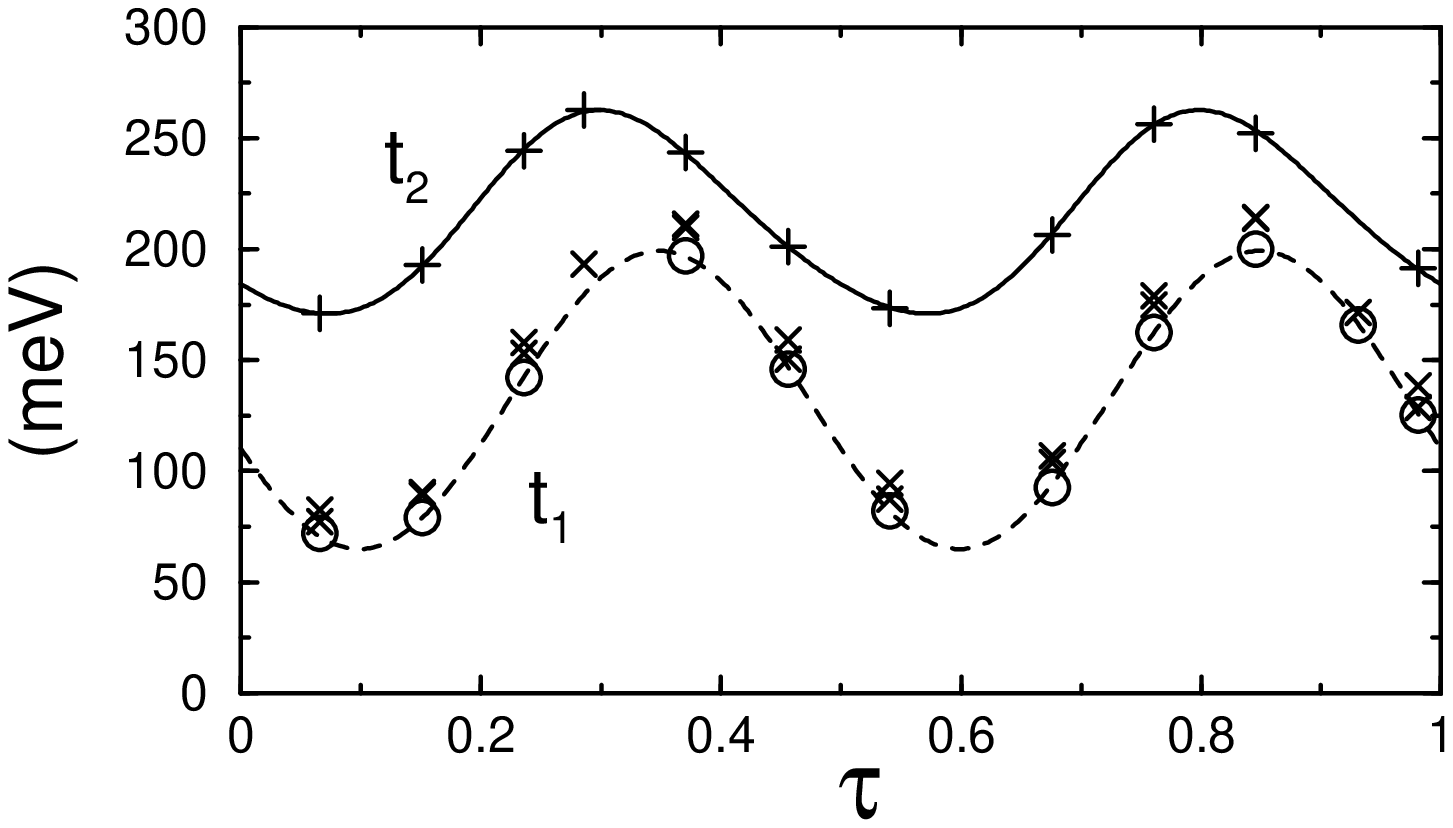}}
\caption{NN and NNN effective hopping integrals as a function of
$\tau$. The circles correspond to the two-coppers fragment
calculations, the cross to the three-coppers calculations and the
lines to the Fourier fits.}
\label{f:t1t2}
\end{figure}

\begin{figure}[h] 
\center
\resizebox{7cm}{!}{\includegraphics{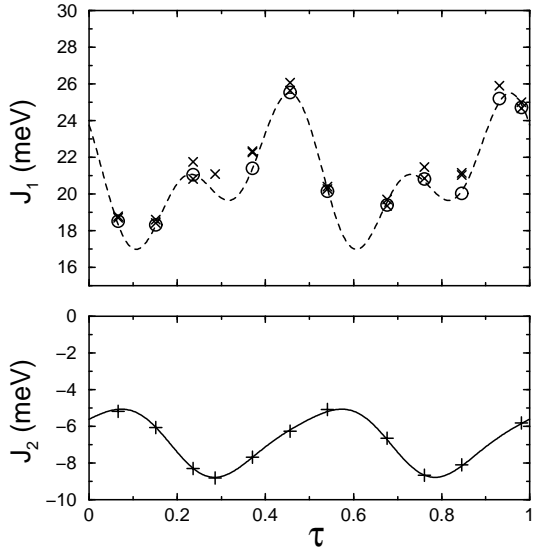}}
\caption{NN and NNN effective exchange integrals as a function of
$\tau$. The circles correspond to the two-coppers fragment
calculations, the cross to the three-coppers calculations and the
lines to the Fourier fits.}
\label{f:J1J2}
\end{figure}

Figure~\ref{f:t1t2} reports the effective hopping integrals while
figure~\ref{f:J1J2} reports the effective exchange integrals. Both are
given as a function of the fourth crystallographic coordinate $\tau$
(defined as in the Hamiltonian expression). 

The NN integrals have been evaluated both from the two-centers and
three-centers fragments. The three-centers fragments used in the
calculations have been chosen in successive positions along the chain,
thus, each NN integral appears in two successive fragments yielding
independent evaluations.  One should first notice that the three
independent evaluations of the the NN hopping, $t_1$, and exchange,
$J_1$, integrals yield the same values, thus validating (i) the second
neighbor $t-J+V$ model used in the present work and (ii) the strong
locality of the interactions. Indeed, whether other interactions or
orbitals would have been of importance, whether the integrals
screening effects would have a larger range than the first
coordination shell of the magnetic atoms, the two-centers and
three-centers calculations would have given different evaluations of
the integrals so that to compensate the inability for the model to
reproduce the low-energy local physics.

As expected, the NNN interactions are of the same order of magnitude
as the NN ones. In fact, the second neighbor hopping integrals are
even nearly twice as large in amplitude as the first neighbor
ones. Indeed, while the NN hole orbitals are nearly orthogonal, the
NNN hole orbitals strongly overlap (see figure~\ref{f:orb}b). Another
important point is the strong modulation of the hopping integrals as a
function of $\tau$. The $t_2$ modulation amplitude is 
$88\ \rm meV$ and represents a $15\%$ standart deviation around an
average amplitude of $214 \rm meV$.

NN and NNN exchange integrals are, as expected, different in nature
with ferromagnetic NN exchange and antiferromagnetic NNN one.  This
is due to the usual well known reasons (i) the direct exchange
integral, which is always ferromagnetic in nature, decreases
exponentially with the $C\!u$--$C\!u$ distance and thus becomes
negligible between NNN centers, (ii) the super-exchange term, mediated
via the bridging oxygen $2p$ orbitals and antiferromagnetic, is
hindered in the NN case due to the destructive interaction of the
bridging orbitals.  It results for the NNN exchange a weak,
antiferromagnetic integral, in agreement with the experimental
findings~\cite{Magn96B}. 
Two points are however worth noticing. \begin{itemize}  \itemsep +0.8ex
\item The experimental evaluation of the intra-dimer exchange $J_d$ is
$-11\,\rm meV$, that is somewhat larger than our findings for NNN
exchange.
\item The computed $J_2$ integral strongly vary along the chain since
its amplitude variation is $55\ \%$ of its mean value $-7\ \rm
meV$. This strong modulation is however not observed in neutron
scattering experiments~\cite{Neut99}.
\end{itemize}
These two important points are discussed  in the following sections.

\section{Super-exchange mechanism through a double bridge}

The DDCI method used in this work is well known to give reliable and
very accurate evaluation of the NN exchange integrals. It has
been used with a lot of success for computing NN exchange integrals in
copper~\cite{DDCIhtc,AbI_La2CuO4}, nickel~\cite{DDCI_Ni},
vanadium~\cite{vana} and other oxides. Indeed, it was  shown that
in order to accurately compute magnetic integrals, the following
effects should be treated~\cite{deLoth,GdCAS,ddci2,Revue}~:
\begin{itemize} \itemsep +0.8ex
\item the multideterminantal and open-shell character of the reference
wave function,
\item the correlation within the magnetic orbital shell, 
\item the screening effects (dynamical polarization) on the different
configurations of the reference wave function, 
\item the effect of the ligand to metal charge transfers
configurations that mediate the interaction, 
\item and the screening effects on the latter configurations.
\end{itemize}
The characteristic of the DDCI method is to treat all these effects in
the case of interactions mediated by at most one bridging entity.

\begin{figure*}[t] 
\center
\resizebox{17cm}{!}{\includegraphics{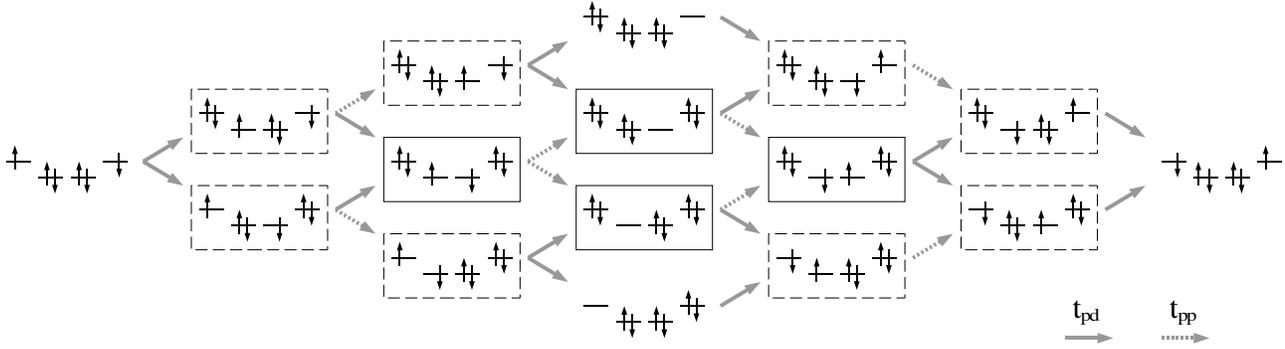}}
\caption{Perturbative mechanism of the NNN effective exchange
integrals mediated through two bridging entities. In each
configuration, the relative position of the orbitals are pictured in
energy scale, that is the two higher orbitals are the magnetic copper
ones and the two lower orbitals are the $2p$ bridging entities
orbitals. The single ligand-to metal charge transfer configurations
are frame with dashed lines and the double transfer ts with solid
lines. The copper to ligand $t_{pd}$ hopping integrals are pictured
with solid arrows while the ligand to ligand hoppings, $t_{pp}$, are
with dashed arrows.}
\label{f:mecaj2}
\end{figure*}

In the present case, the NNN interactions are mediated by {\bf two}
and not one bridging entities, namely the two oxygens between the
first and second copper and the two oxygens between the second and
third copper. It results a perturbative evaluation of the $J_2$
integral within a two bands (magnetic plus bridging entities)
extended Hubbard model of
$$ J_2 = - 4 \frac{t_{pd}^4 t_{pp}^2}{\Delta_1^4 U_d} 
         - 4 \frac{t_{pd}^4 t_{pp}^2}{\Delta_1^4 \Delta_2^\prime} 
         - 16\frac{t_{pd}^4 t_{pp}^2}{\Delta_1^3 \Delta_2 \Delta_2^\prime} 
         - 16\frac{t_{pd}^4 t_{pp}^2}{\Delta_1^2 \Delta_2^2 \Delta_2^\prime} 
$$ where $t_{pd}$ is the hopping integral between the copper magnetic
orbital and the bridging oxygen $2p$ orbitals and $t_{pp}$ is the
hopping between the orbitals of the two bridging entities. $\Delta_1$
is the excitation energy between the reference configurations and the
single ligand-to-metal charge transfer configurations.  $\Delta_2$ and
$\Delta_2^\prime$ are the excitation energies of the double charge
transfers configurations, $\Delta_2$ when the holes are on different
bridging entities and $\Delta_2^\prime$ when they are on the same.
Figure~\ref{f:mecaj2} pictures the different perturbative mechanisms
yielding the above equation.
One sees  that out of the twenty possible
paths, only two of them do not involve double ligand-to-metal charge
transfer configurations. Thus, even-though the double charge transfer
excitations are of higher energy than the single ones (of the order of
twice as large), they are much more numerous and thus should be
responsible for a large part of the $J_2$ effective integral, unlike
the case of single bridged integrals.  One should now remember that
while the DDCI method correctly compute the screening effect of the
ligand-to-metal charge transfer configurations, this is not the case
for the double charge transfer ones~\cite{GdCAS}. It results that the
excitation energies toward the double charge transfer configurations
($\Delta_2$ and $\Delta_2^\prime$) are overestimated and the that thus
the $J_2$ integral is underestimated in the present calculation.
Unfortunately including the screening effects of the double charge
transfers is technically not possible at this time since it would
involve triply excited configurations compared to the reference ones.

It is important to point out that a similar perturbative analysis of
the NN integrals as well as the NNN hopping integral shows that the
contribution of the double charge transfers are negligible in these
cases. Moreover, noticing that the $J_2$ modulations are essentially
due do the modulations of the transfer integrals ($t_{pd}$ and
$t_{pp}$), the relative variations of $J_2$ can be assumed to be
correctly reproduced in the present calculations.

\section{Second neighbor exchange fluctuations along the chains}

Let us now address the question of the NNN exchange fluctuations along
the chain. As previously mentioned the neutron scattering experiments
do not see much fluctuations of the intradimer effective exchange.  In
order to compare our results with the experimental ones one should
determine which among all the NNN exchange integrals along the chain
correspond to an intradimer spin coupling.

In incommensurate systems, the real space analyses present the
drawback to never provide a complete description of the total
system. We thus chose to identify the dimer localization along the
chain not in the real space but as a function of the fourth
crystallographic coordinate $\tau$. Even if such an analysis is not as
intuitive, it presents the advantage to provide a complete description
of the system.

We will thus analyze the system in order to derive a partition into
physically pertinents blocks and to determine (i) the number of
different type of blocks, (ii) their relative arrangement and (iii)
their rate of existence in the incommensurate structure. For this
purpose it is useful to start with a chosen type of block, expected
to be largely represented in the system. In the present case we chose 
the dimeric units. Then, the model Hamiltonian is used to determine
the apparition of this type of blocks as a function of the fourth
component. The analysis of the remaining sites allows to determine the
other types of significant blocks present in the system. Finally, the
relative arrangement of the types of blocks can be studied.

\begin{figure}[h] 
\center
\resizebox{6cm}{!}{\includegraphics{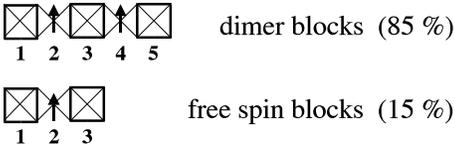}}
\caption{The different types of spin blocks appearing in the
chains. Values in parentheses refer to the percentage of the chain
corresponding to the different types of blocks.}
\label{f:blocs}
\end{figure}

\begin{figure}[h] 
\center
\resizebox{7cm}{!}{\includegraphics{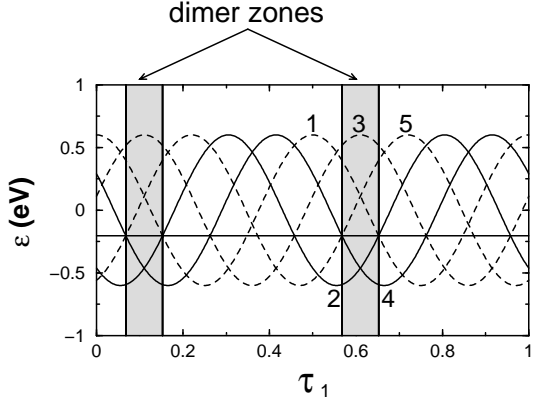}}
\caption{Orbital energy of five consecutive sites as a function of
the fourth coordinate $\tau_1$ of the first one. The horizontal line
represent the Fermi level.}
\label{f:e5}
\end{figure}

\begin{figure}[h] 
\center
\resizebox{7cm}{!}{\includegraphics{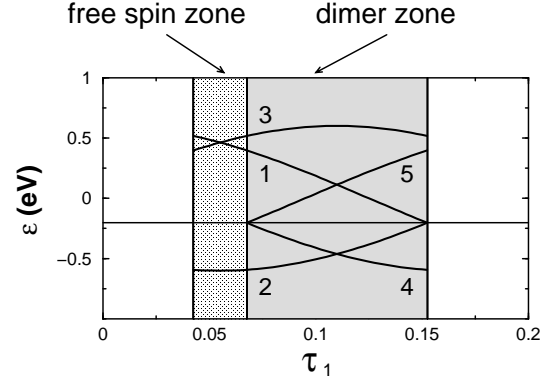}}
\caption{Orbital energy of block sites for the two different types of
block given as a function of the fourth coordinate $\tau_1$ of the
first site of each block. The horizontal line represent the Fermi
level.}
\label{f:e5+e3}
\end{figure}

The dimeric units are composed of five sites, the three
sites of the dimer plus the two ZR singlets of the neighboring sites
(see figure~\ref{f:blocs}). 
Let us use the fourth coordinate of the first site ($\tau_1$) as the
block reference. If one plots the energy of five consecutive sites as
a function of $\tau_1$, a dimeric unit will thus be obtained when the
energy of the second and fourth sites are below the Fermi level, while
the energy of the first, third and fifth sites are above the Fermi
level. Figure~\ref{f:e5} shows the orbital energy curves of five
consecutive sites as a function of $\tau_1$. Let us notice that, since
five sites are represented by a unique value of $\tau$, only one fifth
of the $[0,1]$ range is necessary to represent the whole chain.
The values of $\tau_1$ for which the five consecutive sites form a
dimer are represented in gray. They span two ranges of width
$0.085$. The proportion of the chain occupied by dimers thus
corresponds to five time the above ranges, that is $85\%$ of the
whole chain. 
The number of dimers can then be evaluated to $1.70$
per f.u., to be compared to $1.47$ dimer per f.u. deduced from
magnetic susceptibility measurements~\cite{Magn96B}. Let us now
analyze the composition of the remaining $15\%$ of the chain. For this
purpose we will determine the number of sites separating two
consecutive dimers. If $\tau_1$ is referencing the first dimer (that
is $\tau_1 \in [0.068\, ; \, 0.153] \cup [0.568\, ; \,0.653]$) then
the reference of the second dimer is given by $\tau_1^\prime = \tau_1
+ n(\tau_1) c_{c}/c_{l}$, where $n(\tau_1)$ is the smaller integer
such that $\tau_1^\prime \in [0.068\, ; \, 0.153] \cup [0.568\, ;
\,0.653]$. At this stage three cases can occur. \begin {enumerate}
\item case. $n(\tau_1)<5$. In this case, the two successive dimers overlap,
in other words there exist physical entities larger than the dimers.
\item case. $n(\tau_1)=5$. In this case, the two dimers are strictly
consecutive along the chain. 
\item case. $n(\tau_1)>5$. In this case, the two successive dimers are
separated by one or several other type of blocks of total length~:
$n(\tau_1)-5$.
\end{enumerate}
In the present system, only $n=5$ and $n=8$ occurs. As stated above
the former corresponds to consecutive dimers and the latter to dimers
separated by three sites blocks. As for the dimers, we will reference
these blocks by the fourth coordinate of their first site, that is
$\tau_1 + 5 \, c_{c}/c_{l}$. The three sites blocks thus span two
ranges~: $[0.042 \, ; \, 0.068]$ and $[0.542 \, ; \, 0.568]$.
Figure~\ref{f:e5+e3} represents the three sites blocks and the dimer
ranges for $\tau_1 <0.5$. In each region, the orbital energy of each
site of the block has been represented.
One sees immediately that the three sites blocks are formed by one
spin surrounded by two ZR singlets. Such a configuration can be
associated with free spins since the nearest neighbor spin is two ZR
singlets afar. Let us notice that three times the free spins ranges
yields $0.15$, that is the total missing part of the chain. The number
of free spins can be now easily evaluated to $0.5$ per f.u., to be
compared to $0.55$ free spins per f.u. obtained from the magnetic
susceptibility experiments. 

A further analysis of the figure~\ref{f:e5} shows that the dimers are
arranged in clusters of three or four dimers separated by a free
spin. It can be evaluated that $54\%$ of the dimers form three-dimers
clusters while $46\%$ form four-dimers clusters.

\begin{figure}[h] 
\center
\resizebox{7cm}{!}{\includegraphics{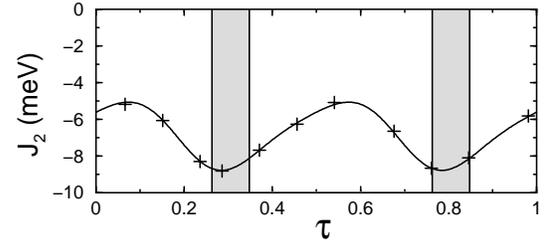}}
\caption{NNN effective exchange integrals as a function of $\tau$. The
regions in gray correspond to intradimer effective exchange
integrals. Let us note that the $\tau$ values in the present figure is
associated with the $\tau_2$ value of the second site of the 5 centers
dimeric units. This results in a shift in the
$\tau$ values compared to figure~\ref{f:e5}.}
\label{f:J2}
\end{figure}

Let us now go back to the intradimer integrals. The regions of $\tau$
corresponding to dimers are reported in gray on figure~\ref{f:J2}. 
The intradimer $J_2$ regions correspond to $J_2$ values between $-8.1$
and $-8.8\,\rm meV$.

Our calculations thus yield values of the intradimer exchange that are
quite homogeneous since their fluctuation range is smaller than $0.7\
\rm meV$, that is less than $8\%$ of their nominal values, in good
agreement with the neutron scattering experiment (the variation range
is within the experimental error bars).  The other interesting point
is that the $J_2$ values corresponding to intradimer exchanges are the
largest ones in absolute values.  The underestimation due to our
method is thus quite limited since our average value is $-8.5\ \rm meV$
to be compared to the experimental values ranging from $11$ and $12\
\rm meV$.

\section{Conclusion}

To summarize the present results, we have determined a second neighbor
$t-J+V$ model for the incommensurate chain subsystem of the
$S\!r_{14}C\!u_{24}O_{41}$ compound. The model parameters have been
determined using accurate ab initio calculations on a series of
embedded clusters along the chain. In order to obtain a complete model
as a function of the incommensurate modulations, the ab initio results
have been extrapolated using a Fourier analysis. The resulting model
is thus independent of any periodic approximation and is 
given as a continuous function of the fourth crystallographic
coordinate $\tau$, which describes the incommensurate modulations along
the chain. 

It is noticeable that, unlike what is currently assumed in the
literature, the various parameters of the model, except for the first
neighbor bi-electronic repulsion, strongly vary as a function of the
structural incommensurate modulations. In fact, these variations are
so large that they determine the physics of the system. In particular,
the orbital energies vary over a surprisingly large range, and thus
dominate the low energy physics through a strong localization of the
electron (resp. holes) over the low (resp. high) energy sites. 

The analysis of the model as a function of the fourth crystallographic
coordinate $\tau$ allowed us to show that at low temperature the chain
ground state can be entirely described only by second-neighbor dimers
and free spins if no hole transfer to the ladder is assumed. It is
worth to point out that our estimation of the free spin number per
formula unit, $0.5$, is under these filling hypothesis in very close
agreement with the experimental value of $0.55$.  Further analysis of
our results showed that the dimers are arranged in clusters of three
or four units separated by a free spin. The intradimer exchanges are
found to be antiferromagnetic, weakly dependent of the dimer position
along the chain and in reasonable agreement with the experimental
values, even if a little underestimated. It is amazing to realize that
while the strong modulation of the model parameters are responsible of
the low energy properties of this system and in particular of the
electron~/~hole localization and the resulting formation of the
dimers, the parameters range actually seen in magnetic experiments
corresponds to a nearly homogeneous dimer system.

Wave functions analysis exhibited holes localization not on the copper
atoms but rather on the surrounding oxygen $2p$ orbitals, thus
confirming the hypothesis of the presence of Zhang-Rice
singlets. Three types of Zhang-Rice singlets can be identified in our
calculations, namely the intra-dimer ones, and two types of
inter-dimer ones~: those neighboring dimers and those neighboring free
spins. This result is to be put in perspective with the copper NMR
experiments~\cite{RMN98} that sees a splitting of the inter-dimer
Zhang-Rice singlets signal at low temperatures. It would be
interesting to quantify the relative weight of the two signals in
order to check our predicted ratio of $3.4$.

{\bf Acknowledgment :} the authors thank Dr. J. Etrillard for
providing us with the neutron crystallographic structures as well as
for helpful discussions, Dr. D. Maynau for providing us with the CASDI
suite of programs, Dr. P. Sciau for introducing us with the
four-dimensional crystallographic conventions. The present
calculations were done using the IDRIS computational center facilities
under project number 1104.

\end{document}